\documentclass[12pt, preprint]{aastex}
\usepackage {psfig, epsfig}

\begin{document}

\title{Far Infrared Observations of Radio Quasars and FR II Radio Galaxies}

\author{Y. Shi\altaffilmark{1}, G. H. Rieke\altaffilmark{1}, D. C. Hines\altaffilmark{2}, G. Neugebauer\altaffilmark{1},
M. Blaylock\altaffilmark{1}, J. Rigby\altaffilmark{1}, E. Egami\altaffilmark{1}, K. D. Gordon\altaffilmark{1}, 
A. Alonso-Herrero\altaffilmark{3}}
\altaffiltext{1}{Steward Observatory, University of Arizona, 933 N Cherry Ave, Tucson, AZ 85721, USA}
\altaffiltext{2}{Steward Observatory, currently at Space Science Institue 4750 Walnut Street, Suite 205,
Boulder, Colorado 80301}
\altaffiltext{3}{Steward Observatory, now at Instituto de Estructura de la Materia, CSIC, Madrid, Spain}

\begin{abstract}
We report MIPS photometry of 20 radio-loud quasars and galaxies at 24 and 70 $\mu$m (and of five at 160 $\mu$m).
We combine this sample with additional sources detected in the far infrared by $IRAS$
and $ISO$ for a total of 47 objects, including 23 steep spectrum
Type I AGNs: radio-loud quasars and broad line radio galaxies; and 24 Type II AGNs:
narrow line and weak line FR II radio galaxies. Of this sample, the far infrared emission of all but 3C 380 appears
to be dominated by emission by dust heated by the AGN and by star
formation. The AGN appears to contribute more than 50\% of the far
infrared luminosity in most of sources. It is also expected that
the material around the nucleus is optically thin in the far infrared. 
Thus, the measurements at these wavelengths can be used to test the orientation-dependent
unification model. As predicted by the model, the behavior of the sources is consistent with the presence
of an obscuring circumnuclear torus; in fact, we find it may still have significant optical depth at 24 $\mu$m.
In addition, as expected for the radio-loud quasars, there is a significant
correlation between the low frequency radio (178 MHz) and the 70 $\mu$m emission, two presumably
isotropic indicators of nuclear activity. This result is consistent with the simple unified scheme.
However, there is a population of radio galaxies that are
underluminous at 70 $\mu$m compared with the radio-loud quasars and hence are a challenge
to the simple unified model.
\end{abstract}


\section{Introduction}
Unification models of active galactic nuclei (AGN) hypothesize that apparently distinct types
of AGN harbor intrinsically similar nuclear engines that may be described by a small number
of fundamental physical parameters.
For example, it is proposed that orientation-dependent obscuration by
a torus surrounding the central supermassive black hole, along with beamed
emission, is responsible for the large diversity in the
appearance of radio-loud AGNs (as reviewed by e.g. Urry \& Padovani 1995).
In the purest form of the model, the torus is always of the same
geometry, so differences among sources arise only due to different inclinations
relative to the observer and different AGN luminosities.
In this paper, we focus on the unification initially proposed 
by Orr \& Browne (1982) and Barthel (1989) for steep-spectrum radio quasars and FR II
radio galaxies (Fanaroff \& Riley 1974).

Recent $Infrared$ $Space$ $Observatory$ ($ISO$)
observations have shown that thermal emission by dust is the dominant
infrared (IR) emission component for most steep-spectrum extragalactic radio sources 
(Haas et al. 1998; Polletta et al. 2000; van Bemmel et al. 2001; Andreani et al. 2002;
Siebenmorgen et al. 2004). However the relative luminosity contribution
from star formation and from the nuclear engine is not clear. Polletta et al. (2000) suggest
the stellar-powered contribution to the IR emission is less than 27\% based
on the relatively warm IR colors indicated by the ratio $F(60~ to~ 200{\mu}m)$/$F(3~ to~ 60{\mu}m)$. 
Although the IR emission from AGNs is generally
warmer than that from star-forming regions, the IR color of AGNs can be affected by
extinction, especially for mis-aligned sources, and the IR colors $F(60{\mu}m)$/$F(25{\mu}m)$ 
of AGNs and star-forming regions have a
large range (Kewley et al. 2001). Therefore, the color-based arguments may not
be rigorous. Theoretically, early models for AGNs (Pier \& Krolik
1993; Granato \& Danese 1994) underestimate the far-IR (FIR) emission from heated dust,
which lead Rowan-Robinson (2000) to suggest a star formation origin for
most of the FIR emission, consistent with results of recent radiative transfer models (van Bemmel et al. 2003). 
Recently, Nenkova et al. (2002) and Siebenmorgen et al. (2004) 
suggest that the nuclear engine can heat the dust to large distances from the galaxy center
and therefore that star formation need play only a small role in the dust heating.

In this paper, we combine IR measurements of radio-loud AGN obtained with the Multiband Imaging 
Photometer for $Spitzer$ (MIPS) (Rieke et al. 2004)
and previous data from the $Infrared$ $Astronomical$ $Satellite$ ($IRAS$) and
$ISO$. We use these observations to show that the IR emission is usually dominated
by radiation from dust heated by the AGN. Thus we can use the far infrared properties of
these sources to gain new insights to unification models.

\section{Observations}
\subsection{The Sample and Extended Sample}
We have used MIPS to observe twenty FR II radio galaxies and steep spectrum radio quasars
listed in Tables 1 and 2. The sample is drawn from Neugebauer et al. (1986) and Golombek et al.
(1988), with priority on objects at z $<$ 0.4 and with
good $Hubble$ $Space$ $Telescope$ ($HST$) images.  To
complement these measurements, we defined an extended sample, consisting of the 27 additional 3CR sources
listed in Table 3, all of which are detected by $IRAS$ or $ISO$
at least at two wavelengths between 10
$\mu$m and 200 $\mu$m.
The $IRAS$ data are from the NASA/IPAC Extragalactic Database (NED)
and the $ISO$ data are from Haas et al. (2004), van Bemmel et al. (2001) and NED. The
uncertainties in the $ISO$ data from Haas et al. (2004) are quoted as 10-30\%;
in this paper we adopt 30\% uniformly. Except for Pic A and 3C 218 
the sources belong to the parent 3CR sample. For this study, we divide the sample into two
subsamples as described in Section 3.1: Type I sources including radio quasars and broad line
radio galaxies (BLRG) and Type II sources composed of narrow line
radio galaxies (NLRG) and weak line radio galaxies (WLRG).

To investigate possible selection biases for our sample, in Figure 1
we plot the distribution of all the sources in the plane of 178 MHz
flux density versus redshift (top box) and 178 MHz flux density versus spectral
index $\alpha$ (bottom box; $\alpha$ is defined as
$f_{\nu}{\propto}~ \nu{^{-\alpha}}$ between 38 MHz and 750 MHz). The
top box indicates for our sample that the 178 MHz flux density spans a range from 6 Jy to
10$^{4}$ Jy and the redshift a range from 0.03 to 1.4. From this
figure, we see that the flux density distributions of Type I and II sources
are similar. The Kolmogorov-Smirnov (K-S) test indicates a
probability of 82\% that the two subsamples have the same distributions.
However, the sample contains many more high redshift (z$>$0.3) Type I sources,
mainly radio quasars, than high-redshift type II sources. The bottom box indicates
that the spectral indices of the two sub-samples are mainly in the range
between 0.5 and 1.2. The K-S test gives a probability of 89\%
that both types have the same distributions of spectral index.
Based on figure 1, we find that our sample is not
biased in AGN characteristics except that there are relatively more Type I sources
at high redshift and thus high luminosity.

\subsection{Infrared Data}
Our observations were made with the standard MIPS small field photometry mode.
The effective integration time for each source is listed in Table 1. The data
were reduced with the MIPS instrument team Data Analysis Tool (DAT) version 2.73 (Gordon et al. 2004a, b).

To obtain the 24 $\mu$m fluxes, we performed aperture photometry using
the IDL-based image processing package IDP3 (Schneider \& Stobie 2002). 
We set the radius of the object aperture to 6 pixels (15.0$''$)
and we measured the background to be the median flux in a 5 pixel (12.5$''$) wide
annulus, at a central radius 11 pixels (27.4$''$) from the centroid of the
target. Contamination from other sources within the sky annulus was
masked by hand. We multiplied by a factor of 1.146 to correct the measured values for the
portion of the point spread function lying outside the source aperture.
The pixel-to-pixel fluctuations in the background region were used to calculate the 1-sigma
uncertainties. The results are listed in Table 1. Most of the sources are
detected at nominally very high signal-to-noise ratios. 
In addition, there is a calibration uncertainty of up to 10\%.

The 70 and 160 $\mu$m photometry was reduced in a similar manner. Additional processing steps, such as 
bad column removal and time filtering, were performed for 70 $\mu$m data. 
The radius of the object aperture at both wavelengths was 3.0 pixels (30$''$ at 70 $\mu$m and 48$''$ at 160 $\mu$m) 
and the inner and outer radii of
the background region were 4 pixels and 8 pixels (40$''$ and 80$''$ at 
70 $\mu$m; 64$''$ and 128$''$ at 160 $\mu$m), respectively.
Aperture correction factors of 1.3 and 1.5 were applied to the 
70 and 160 $\mu$m measurements respectively.
The results of the 70 $\mu$m photometry are also listed in Table 1.
The five sources detected (in a 3-pixel aperture)
at 160 $\mu$m are listed in the footnotes to Table 1.
There is a calibration uncertainty of up to 20\% at both 70 and 160 $\mu$m.

To combine the extended sample with the MIPS sample, we need to
interpolate the $IRAS$ and $ISO$ data to calculate the fluxes at the
MIPS bands. We use the data points at the two wavelengths closest to the
MIPS band for interpolation. Multiple observations at one wavelength are averaged. We
assume power law SEDs to
interpolate to the MIPS band and also to propagate the errors. The
resulting fluxes are listed in Table 3.
Additional errors can arise from the difference between
the power law and the real SEDs. By comparing interpolations using
a suite of blackbody (T $\ge$ 40 K) and power law SEDs, we estimate
that such errors should be $<$ 30\%.

\subsection{HST Images}

We retrieved archived $HST$ data for all sources in our sample, except
for 3C 218 which has not been observed. The images were
taken with the Wide Field and Planetary Camera 2 (WFPC2), generally with
the F555W or F702W filters. The data were processed through the PODPS
(Post Observation Data Processing System) pipeline to remove bias
and flat-field artifacts (Biretta et al. 1995). Individual exposures in an
observation were combined to remove cosmic ray events. For observations
with only one exposure, the cosmic rays were masked by interpolating over them.

\subsection{Chandra Data}
We retrieved data from the Chandra archive for 19 sources observed with the Advanced CCD
Imaging Spectrometer (ACIS). Data
reduction and analysis were performed using CIAO 3.0.2. The level 2
event file was made from the level 1 event file after correction for gain,
aspect and charge transfer inefficiency, and also after PHA
randomization and destreaking. We used the light curve to remove
solar flares. Inspection of the images showed that
there was severe pileup in the central region of 3C 390.3 where a hole without counts was present. 
The nuclei of 3C 348 and 3C 173.1 had too few counts for our analysis 
and they were dropped from the X-ray sample. For the remaining sources, we extracted the
spectra of the nuclear region with an aperture diameter of 2.5$''$ around the peak
pixel and then fitted the rest-frame 2-10 kev spectra using an absorbed
power law. Finally we computed the absorption-corrected hard-X-ray
fluxes, which are listed in Column (10) of Table 2 and Column (13) of Table 3.
The photon spectral indices are listed in Column (9) of Table 2 and Column (12) of Table 3.

\section{Analysis}

\subsection{Division of the sample into quasars and galaxies}

We classified the sources as Type I (radio quasar or BLRG) or Type II (NLRG
or WLRG) on a consistent basis. Emission line types were obtained from the literature. The
division between radio quasars and BLRG was based on
absolute B magnitude. We retrieved the apparent Johnson B magnitude
from NED, originally from Sandage et al. (1965), Smith \&
Heckman (1989) and de Vaucouleurs et al. (1991). For those sources
without available blue magnitudes, we took V magnitudes from Spinrad et al. (1985) and
assumed $<$B$-$V$>$=0.3 mag. The conversion formula
from the apparent magnitude to absolute magnitude is: 
\begin{equation}
M=m-5log(D_{L})+5-K(z)-A_m
\end{equation}
where $M$, $m$ are absolute magnitude and apparent magnitude, $D_{L}$
is the luminosity distance, $K(z)$ is the K-correction at the redshift $z$
and $A$ is the Galactic extinction. 
$A_m$ is computed using NED. The majority of our sample are in
luminous elliptical galaxies; we obtained K-corrections for such galaxies 
from Pence (1976). We assumed H$_{0}$ $=$
75km s$^{-1}$ Mpc$^{-1}$, $\Omega_{\rm M}$=0.3 and $\Omega_{\Lambda}$=0.7. 
The resultant absolute blue magnitudes are listed in Column (2) of
Table 2 and Table 3. Radio quasars have power law SEDs, typically with 
a spectral index of 0.5. Therefore, we also calculated the absolute 
B magnitude for all sources using the
K-correction based on a power law SED with a spectral index of 0.5 

We  defined as quasars those broad line sources with an absolute B magnitude $<$ -23 mag and as BLRGs those with
fainter absolute B magnitudes. Only four sources have ambiguous classifications. 
3C 318, 3C 325, 3C 332 and 3C 382 were classified as radio quasars
using the K-correction based on the SED of normal elliptical galaxies, 
while they were radio galaxies if we assumed their optical SEDs were power laws.  
The remaining galaxies without broad lines were classified
as NLRG or WLRG according to the strength of their emission lines.
Column (3) of Table 2 and Table 3 list the results. There are 17 radio quasars, 
6 BLRG, 20 NLRG and 4 WLRG. We emphasize that the ambiguous classifications of a few
sources do not affect the results of this paper.
    
\subsection{Spectral Energy Distribution}

For each source, we used NED to compile the SED from radio to millimeter, as presented in Figure 2.
The radio and millimeter-wave SEDs of these sources are smoothly steepening power laws
associated with synchrotron emission. This behavior provides the basis for us to
extrapolate to determine the non-thermal contribution to
the IR emission. This process requires a synchrotron
radiation model and good high-frequency data, since the break frequency of the synchrotron emission
occurs at $\sim$ 1Ghz (Polletta et al. 2000). We used a parabola approximation (Andreani et al.
2002) to fit the synchrotron emission. The spectrum in this
model is given by
\begin{equation}
logF_{\nu}=C+\frac{1}{2A}(log\nu-log\nu_{t})^{2}
\end{equation}
where $C$ is a constant, $\nu_{t}$ characterizes the spectral break 
(i.e., where the nominal optical depth is unity) and
$\frac{1}{A}=2\alpha_{1}-\alpha_{2}$, where $\alpha_{1}$ and
$\alpha_{2}$ are the spectral indices above and below the break,
respectively. For each source, we adjusted $C$, $A$ and $\nu_{t}$ to fit all data at
wavelengths longer than 1 mm. Figure 2 shows the results; the extrapolation
is well constrained since most sources have high-frequency 
data and the model can usually fit the data well. 
The IR emission is generally well above the non-thermal
extrapolation. Five sources, 3C 218, 3C 315, 3C 348, 3C 349
and 3C 380, seem to be exceptions: they have IR fluxes below the non-thermal extrapolation at 
one or more bands. Except for 3C 380, the remaining sources have no
sub-millimeter observations and thus their IR emission can be thermal emission by dust 
if the radio spectrum turns over at very high frequencies. Since their 24 $\mu$m to 70 $\mu$m IR color is redder than 
the non-thermal continuum, we argue for the thermal nature of the IR emission for these four sources. This hypothesis
is confirmed by a mid-IR spectrum of 3C 218 that shows PAH emission (our unpublished data).
The IR flux of 3C 380 is along the non-thermal extrapolation as shown in the Figure 2; Bloom et al. (1994) find that the radio spectrum is flat
from around 1 cm up to 1100 $\mu$m, implying that the non-thermal contribution is significant at infrared wavelengths. The 5 GHz radio emission
mainly comes from a region less than 2 arcsec across, as shown by comparing $F_{\rm 5t}$ and $F_{\rm 5c}$ in Table 3.
VLBI observations of the milliarcsecond-scale structure of this compact core reveal a one-sided core-jet (Wilkinson et al. 1990; 
Polatidis \& Wilkinson 1998). The total flux of all compact components of the core-jet is 2-5 Jy at 5 GHz, contributing  30-60\% of the total flux. 
This radio core with flat spectrum may make a substantial non-thermal contribution to the infrared emission of 3C 380. Given the ambiguities in the origin
of the infrared emission, we do not use this source in the following analysis.

We also evaluated contamination from stellar light in the host galaxy.
The stellar light in radio quasars should be negligible. To estimate
the stellar contribution from the other host galaxies, we used the spectral template for a normal elliptical
galaxy over the spectral range from 1400 $\AA$ to 2.75 $\mu$m (M. Rieke, private communication).
The template was shifted by (1+$z$) and then normalized at
4405 $\AA$ to the apparent B magnitude. The dotted line in
Figure 2 shows the stellar emission of the host galaxy; the contribution
in the mid-IR (MIR) and FIR wavelength ranges
will be negligible.

\indent We conclude that the IR emission of
most of our sample of radio-loud galaxies and quasars is mainly thermal IR
emission by dust, except for 3C 380.

\section{Discussion}

\subsection{Energy Source for IR Emission}

We will use several methods to assess the relative contributions of 
forming stars and the AGN in heating the dust in our sample of radio-loud AGNs. 
We find that star formation provides less than 50\% of the IR luminosity in most sources.

\subsubsection{Extension of the IR Emission Region}
The MIPS 24 $\mu$m images do not resolve the sources. The beam diameter of $\sim$ 6$''$
corresponds to $\sim$10 kpc at a redshift of 0.1. We convolved the HST
image with the PSF of MIPS at 24 $\mu$m created by STINYTIM and
found that the 24 $\mu$m image should show some structure for
low-redshift (z $<$ 0.3) sources if the IR emission is from the entire
host galaxy. Thus, the IR emission is constrained to the
central region.

\subsubsection{Emission Line Ratio and IR Color}

Baldwin, Phillips, \& Terlevich (1981), Osterbrock \& de
Robertis (1985) and Veilleux \& Osterbrock (1987) have developed
spectral diagnostics to classify emission-line galaxies and
determine the dominant energy source, star formation or an active
nucleus. They compare two emission-line ratios combining high and low excitation
lines, such as [O III] $\lambda$5007/H$\beta$, [N II] $\lambda$6583/H$\alpha$, or [S II]
($\lambda$6716+ $\lambda$6731)/H$\alpha$, in a two dimensional classification diagram.

The dust heated by an AGN is usually warmer than that heated by star
formation and thus the IR color can be used instead of one emission line
ratio for classification diagnostics (Kewley et al. 2001). 
We use [O III] $\lambda$5007/H$\beta$ and IR
color to diagnose the dominant source of energy for the IR emission. The
data for the ratio [O III] $\lambda$5007/H$\beta$ are collected from
the literature and the extinction corrections follow
Veilleux \& Osterbrock (1987). Our extinction estimates assume an intrinsic ${I({\rm
H}{\alpha})}/{I({\rm H}{\beta})}=3.1$ (Kewley et al. 2001).
For several sources, we use the ratio of equivalent widths of the
emission lines because the flux ratio is not available. The extinction 
corrections are small in virtually all cases. Some
sources could not be corrected for extinction because there
are no measurements of H$\alpha$. The errors introduced by uncorrected
line strengths should be negligible, given the small corrections for the
other sources. 

Figure 3 shows the distribution of 22 sources in the
emission-line-ratio versus IR-color plot.
The IR colors are defined as the ratios of 60 $\mu$m
flux densities to 24 or 25 $\mu$m ones. For sources without 60 $\mu$m measurements,
the flux densities were obtained by interpolation as in Section
2.2. A large [O III] $\lambda$5007/H$\beta$ ratio
and warm IR color indicate significant AGN activity, i.e., the upper left region of Figure 3. The
lower right part of the figure is the locus of star formation. We
use the hypothetical mixing line (Kewley et al. 2001) in Figure 3 to estimate the
percentage contribution of star formation. Most of our sources lie close to the region
delinated by confirmed AGNs (filled dots), and for them we conclude that 
the AGN contributes $>$50\% of the total power.
However, for a few sources, it appears that the contribution from star formation can 
be dominant, much higher than the estimation made by IR color alone
(Polletta et al. 2000). 

In Section 4.2.1, we show that the circumnuclear torus may have significant optical depth at 24 $\mu$m
and so the IR color of Type II AGN may be reddened through obscuration by the torus.  Thus, 
Figure 3 may underestimate the fraction of the AGN contribution
for Type II AGN. 

\subsubsection{Relation of IR emission to the Central Hard X-ray Emission}

Since the thermal IR emission appears usually to be powered by the nuclear engine, under
the unified model we might expect
a relation between IR luminosity and nuclear
luminosity at other wavelengths. 
Such a relation might be hidden at wavelengths with strong obscuration by the circumnuclear
torus. Hard X-rays are largely
immune from such effects. The
weakness of the correlation between hard-X-ray flux and core
radio flux at 5 GHz as shown in Table 4 indicates that the hard X-ray emission for our sources
is mainly from the accretion disk and other processes around the central
engine, not from the beamed emission.
Hard X-rays from the accretion disk are believed to provide a
reasonably isotropic estimation of the nuclear luminosity.

Figure 4 compares the rest-frame 2-10 kev
absorption- and K-corrected X-ray flux in the central 2.5$''$ region
and the 70 $\mu$m and 24 $\mu$m K-corrected flux densities. 
Because of the inhomogeneity of our sample, 
we compare K-corrected flux densities
rather than luminosities. The spectral indices of the power laws assumed
for the K-corrections of the
X-ray measurements are listed in Column (9) of Table 2 and Column (12) of Table 3. 
The K-corrections for the IR bands were calculated assuming a power law
SED with spectral index of 1. The Kendall Tau test on these data (see Table 4)
yields $S$ values of 0.05 and 0.09 respectively at 24 and 70$\mu$m, showing only 5\% and 9\%
probabilities respectively that the measurements are uncorrelated. 

Except for 3C 380, all sources in Figure 4 have negligible non-thermal IR emission. 3C 321 is well above the correlation in Figure 4
and may have significant star formation activity. Additional evidence for active star formation in 3C 321 includes: 3C 321 
contains two close nuclei (Roche \& Eales 2000) and large amounts of star formation may be triggered 
by mergers (Sanders et al. 1988); Tadhunter, Dickson, \& Shaw (1996) find that 
the AGN fraction can be as low as 26\%  of the UV continuum by fitting the observed spectra of 3C 321.

The correlation indicates that the thermal IR emission is
associated primarily with the nuclear output, not with processes such as star
formation that would operate independently of nuclear power. Thus, the IR
emission is largely the emission of dust heated by the central engine.
Because of the lack of suitable X-ray detections, the radio galaxies are
relatively poorly represented in Figure 4, but there is no reason to expect
them to behave differently from the other sources.

For the following discussion, we fit the IR/X-ray correlation in luminosity rather than flux density, 
since converting to this form and fitting the slope reduces the uncertainties due to our rather
crude K-corrections. Excluding 3C 380 and 3C 321, 
the fits are given by $Log(L_{70})=1.68+0.89Log(L_{\rm x})$ and
$Log(L_{24})=1.21+0.93Log(L_{\rm x})$. The dispersion of the
relation is given by the relative standard deviation defined by
$<((L_{\rm obs}-L_{\rm theo})/L_{\rm theo})^2>^{0.5}$, where $L_{\rm obs}$ is
the observed IR luminosity and $L_{\rm theo}$ is the theoretical IR
luminosity heated by the central black hole, estimated from the fits
discussed above. The result is 1.4 and 0.7 for the 70 $\mu$m and 24
$\mu$m relation respectively. We can derive a rough upper limit to the
contribution of star formation by assuming that all of the scatter arises from
this process. The dispersion values imply that the contribution 
from star formation is smaller than 60\% and 40\% at 70 and 24 $\mu$m rest-frame wavelengths, respectively. 
This result is consistent with
the conclusion from our emission-line-ratio vs. IR color analysis.

\subsection{Tests of Unification Models}

Since most of the IR emission from these sources seems to be associated
with the AGN, we can use the IR data to probe unification models. We will
describe two types of test: 1.) to see if the distribution of physical parameters
is consistent with an obscuring circumnuclear torus and beamed emission by a jet,
observed over a range of viewing angles; 
and 2.) to check if the relations between selected physical parameters are the same for
the radio quasars and galaxies.

\subsubsection{IR Color vs. $R$ Parameter}

Because of beaming, the emission of the radio core should be
orientation-dependent. Thus, the radio compactness, the $R$ parameter, can be used to
indicate the viewing angle with respect to the orientation of the radio jet (Orr \& Browne 1982).
Here we define $R=F_{\rm 5c}/F_{\rm 5t}$, where $F_{\rm 5c}$ and $F_{\rm 5t}$ are
the core and total radio fluxes at 5 GHz, respectively.
$R=1$ corresponds to the direction along the radio jet
and the smaller $R$, the larger the viewing
angle (we assume the torus is perpendicular to the jet, so $R$ is taken as
an indication of the viewing angle relative to the torus as well as the jet). 
We obtained from NED the total 5 GHz radio emission listed in Column (5) of
Table 2 and Table 3, and from the literature the radio core emission listed in Column (6)
of Table 2 and Table 3. We plot the IR color defined by
$F_{70}$/$F_{24}$ as a function of the $R$ parameter in Figure
5. In this figure, we do not include sources with upper limits
to their IR fluxes, nor the sources with possibly significant non-thermal components or large star formation
contributions ($\sim$80$\%$ in Figure 3). 

We find a trend that redder IR
color corresponds to a smaller $R$ parameter and different
emission-line types fall in different regions in the plot,
qualitatively consistent with the prediction of the unification
model. The overlapping between different emission-line types of
source in the plot is possibly because the $R$ parameter
depends on other characteristics beside the viewing angle (Orr \& Browne 1982), 
such as the core Lorentz factor,
and because intrinsic dispersion in the radio core and lobe emission
causes dispersion in the $R$ parameter (Lonsdale \& Barthel 1987). 
The least square fit to the relation in Figure 5 is given by
$F_{70}/F_{24}=0.35-0.74Log(R)$. The corresponding Kendall-Tau test 
result given in Table 4 indicates a 96\% probability of a significant correlation. 
The observed range of IR color of a factor of $\sim$ 2.5 indicates 
that optical depth effects are still significant at 24 $\mu$m in the typical torus in an FR II radio galaxy. 
In contrast, Heckman et al. (1994) compare 60 $\mu$m fluxes between narrow line AGN and broad 
line AGN and found that they were almost the same,
suggesting the torus is optically thin at 60 $\mu$m. 

For 3C 405, with the smallest $R$ parameter
in our sample, the absorbed power law 
fit to the hard X-ray spectrum shows the hydrogen column density is $N_{HI} = 10^{23} 
cm^{-2}$. Using $A_{V}=0.62*10^{-21}N_{HI}$ (Savage \& Mathis 1979) where $A_{V}$ 
is the visual extinction, we have $A_{V}=62$. Taking $A_{V}/A_{70}= 364$, $A_{70} = 0.17$,
indicating the extinction is small at 70 $\mu$m even when we view the AGN though the torus, in agreement
with our conclusion from the color behavior. 

However, the extinctions at 24 $\mu$m and 70 $\mu$m should not be very different; $A_{24}/A_{70}$ is 
expected to be around 7. Therefore, to explain the systematic change
in optical depth between these two wavelengths may require a specific type of torus model.
We hypothesize that there is a temperature gradient along the radial direction 
of the torus; the inner part of the toroidal disk is closer to the 
central heating source and hence is hotter than the outer region. To maintain 
a gradient may require that the outer parts of the disk are heated by energy reradiated
from the inner parts, that the disk is warped or flared, or that it has a porous (or cloudy) structure. 
Such a temperature gradient may explain the non-single-blackbody IR SED of the AGN as shown in Figure 2. 
As a result, the 24 $\mu$m flux is predominatedly emitted by the dust in the inner torus, while 
70 $\mu$m emission is mainly from the outer region. If the density of the toroidal disk drops with
distance from the central region, then the optical depth to the 24$\mu$m emission from the 
inner region could be large while 
it is still small to a significant portion of the 70 $\mu$m emission from the outer region.

As can be seen in the diagnostic diagram in Figure 3,
the reddening due to an edge-on accretion torus could drive
a pure AGN spectrum toward the star-forming region.
These results indicate that the FIR color is not a completely reliable indicator of star
formation activity; an AGN with a large torus inclination angle can also
have a red FIR color. Two-dimensional diagrams, such as
IR color vs. emission-line-ratio may be useful to indicate the level of star formation activity, though
they are still not infallible.

Recently, Whysong \& Antonucci (2004) argue that the lack of 12$\mu$m emission from some FR I AGNs 
indicates that they lack a torus. However, if the torus has significant optical depth at 24 $\mu$m, the 12 $\mu$m emission 
will be strongly suppressed and thus cannot be used to estimate the intrinsic torus emission.

\subsubsection{An Indication of the Torus Opening Angle}

We return to Figure 4 to compare the IR and X-ray emission for quasars and radio galaxies.
Although the data points in Figure 4 are mainly for radio quasars,
the various kinds of radio galaxy follow the relation well.
Since the IR emission is the reprocessing of the emission from the
accretion disk, especially optical and UV photons, the IR emission
is determined under the unified model by the central intrinsic emission and the torus
structure. The torus structure must be very complicated. However, for simplicity,
we assume it is described only by the opening angle and that the light blocked by the torus
is reprocessed completely to IR emission, otherwise it can
escape completely. Moreover, we assume the emission of the torus
is isotropic at 70 $\mu$m (we also show the behavior at 24 $\mu$m  for comparison).
If the torus opening angle is constant, then the
IR emission should be proportional to the X-ray emission. The least square fit 
indicates the slope is 0.89
and 0.94 for 70 $\mu$m and 24 $\mu$m, respectively. Neither value is significantly
different from unity. Based on the above
simple model, such a slope suggests that the torus opening
angle does not change with the activity level of the central engine.
The radio power of our sources is greater than 10$^{26.5}$ W Hz$^{-1}$
at 178 MHz, so this result is consistent with the invariance of the opening angle based on
the quasar fraction (Willott et al. 2000). The result also indicates that most
radio galaxies and radio quasars possess a torus with similar
structure, supporting the unification model.

\subsubsection{The Relation of IR emission and 178 MHz radio emission}

The 70 $\mu$m emission and 178 MHz radio
emission should also both be isotropic, so we can use them to probe unification models for 
different types of radio source. The relation of the K-corrected
70 $\mu$m flux density with the K-corrected 178 MHz flux density 
is shown in Figure 6. For the radio,
we base K-corrections on a power law SED with spectral index of 0.8, while we
used a spectral index of 1 at 70$\mu$m. The relation between 24 $\mu$m
and 178 MHz emission is not shown because of the many upper limits to the 24
$\mu$m flux. Because the 178 MHz flux density is a defining parameter for the sample, 
Figure 6 includes all the radio galaxy members (in comparison, for example, with
Figure 4 which includes only a fraction). The figure shows the distributions of radio quasars and radio
galaxies are different. Radio quasars except for 3C 380 are
constrained in a tight correlation with an $S$ value of $3\times10^{-3}$
by the Kendall Tau test. The solid line is the least 
square fit to the radio quasars excluding 3C 380 (See Section 3.2); the dashed lines are the 3-sigma bounds to the
relation indicated by the solid line.

Again, 3C 321 is above the 3-sigma bounds. Its behavior may be due to powerful star formation activity triggered by a merger (see Section 4.1.3).
Three additional galaxies close to the 3-sigma high limit (3C 293, 3C 403 and 3C 459) may also have excess IR emission powered by star formation.
Nearly one-third of the radio
galaxies are below the region between the two dashed lines. Except for 3C 405, PicA, 3C 218 and 3C 348, the
FIR underluminous sources do not show
larger radio fluxes compared with radio quasars, so we cannot attribute the behavior in Figure 6 to selection bias.
Since these galaxies are underluminous at 70 $\mu$m, we cannot account for
their behavior by an additional FIR source such as star formation. We conclude that the simple unification model works
less well for the galaxies in the sample than for the quasars; it appears to predict
the properties of some galaxies but not others.  

We have examined the FIR underluminous galaxies individually to see if we can understand why
they depart from the simple unification model prediction. We look for explanations
in terms of their having a weaker nuclear
engine, since the FIR emission is a good indicator of the
nuclear accretion power, or of their having higher radio
emission because of a special environment where radio
emission is easily generated (Barthel et al. 1996; Haas et al. 2004).

3C 388: Based on the relation of the optical core luminosity and
radio core emission, Chiaberge et al. (2000) argue that 3C 388
has a FR I nucleus, i.e., weaker than for classic FR II galaxies.

3C 348: The radio morphology of 3C 348 is neither like FR I nor FR II. It is better classified as an intermediate
type radio source. Thus, we may speculate 3C 348 has a relatively weaker nucleus with respect to
normal FR II galaxies.

We have not found a good explanation for the behavior of the remaining galaxies. Their environments
are not different from the normal FIR luminous sources. Barthel et al. (1996) 
find that sources in X-ray clusters show higher radio-to-infrared 
ratios than non-cluster galaxies. This may be due to the upper-limit measurements of IR fluxes by IRAS, for example, 3C 61.1 and 3C 315 are now
detected by MIPS and their ratios can reach 2.9 and 2.75 respectively and therefore only one source in the sample of cluster galaxies 
shows a significantly higher ratio than for field galaxies. No direct evidence indicates that the remaining FIR underluminous sources have weaker nuclei.
It may be necessary to modify the simple unification model to explain them. For example, there may be a large range
of optical depths at 70$\mu$m; the weakly emitting galaxies could then be significantly obscured.

\subsection{Compact Steep Spectrum Sources}

Compact Steep Spectrum (CSS) and Gigahertz-Peaked Spectrum (GPS) radio sources have similar radio
outputs to other radio sources, but are substantially more compact, 1 to 20kpc. As discussed by O'Dea
(1998), IR observations can distinguish a number of hypotheses regarding these objects, most notably
whether they are {\it frustrated} large-scale sources due to a dense interstellar medium that absorbs the
energy of the radio-emitting jets. This absorbed energy would be expected to emerge in the FIR. A
number of searches have failed to find any effect of the type predicted by the frustrated source model
(Heckman 1994; Fanti et al. 2000).  Because the FIR detection limits available to these studies were
generally inadequate for measurement of single galaxies, they co-added signals from many galaxies to
achieve an average emission level. The levels from the CSS/GPS and control samples are similar, although
the significance level of the comparisons is only modest. The comparisons are also subject to uncertainty
due to the necessity of stacking detections for higher statistical significance: a few extreme sources can
dominate the results. Another probe of these sources is to determine the absorption and hence the density of the surrounding
interstellar medium. Pihlstr\"om et al. (2003) report HI observations that show an increase in ISM density
with decreasing radio size. This behavior is qualitatively consistent with the frustrated source model,
but Pihlstr\"om et al. (2003) show that the ISM density is probably inadequate to confine the jets, unless the
H$_2$/HI ratio is much higher than normal. Fanti et al. (2000) give additional
arguments against the frustrated source model. Our data permit another test of the frustrated source
model, based on IR data with enough sensitivity to detect individual galaxies.

We identify six CSS sources whose radio sizes are smaller than 20 kpc in our sample. The comparision sample
consists of large scale sources in the redshift range of the CSS sample, from 0.27 to 1.4. Figure 7
plots the linear radio size
and IR luminosity. The radio size ranges from 2 kpc to 400 kpc and IR luminosity is
from 10$^{10}$ L$_{\odot}$ to
10$^{14}$ L$_{\odot}$. No correlation is indicated. We conclude that the CSS sources do not show
significantly different FIR output from the large scale radio sources, consistent
with the result of $IRAS$ (Heckman et al. 1994)  and $ISO$ (Fanti et al. 2000). Hes, Barthel, \& Hoekstra (1995) find that CSS sources
are IR-brighter. Their result may be caused by the selection bias that CSS sources in their sample are at high-redshift and
thus high luminosity. In addition, the lack of an additional IR
luminosity component argues against the proposal that these galaxies might be
sites of strong star formation induced by the
interaction between a jet and ambient matter (Baker et al. 2002).

\section{Conclusions}
In this paper, we present MIPS observations of steep spectrum radio quasars and FR II
radio galaxies, and combine them with the sample detected by $IRAS$ and
$ISO$. We discuss the nature of the IR emission in this combined sample and its implications
for the unification model. The main results are:

(1) The IR emission of most sources is thermal. The thermal IR
emission of most sources is dominated by dust heated by
the AGN.

(2) The simple unification model predicts the properties of the
radio quasars and some radio galaxies well. However the properties of some FIR underluminous
galaxies may be inconsistent with a pure unification model in which there are no
orientation-independent intrinsic differences among these sources.

(3) The behavior of the 70$\mu$m/24$\mu$m color with radio compactness is consistent with the suggestion that a
torus that has significant optical depth at 24 $\mu$m surrounds the nuclei of the FR II radio galaxies.

(4) The CSS sources do not show additional IR emission, arguing
against the proposals that these sources reside in extremely dense regions or that they are
associated with strong star-forming activity.

\acknowledgments
We thank John Moustakas for helpful discussions and the anonymous referee for detailed comments. This research has made 
use of the NASA/IPAC Extragalactic Database (NED) which is operated by the Jet
Propulsion Laboratory, California Institute of Technology, under contract 
with the National Aeronautics and Space Administration. This work was supported by NASA 
through Contract Number 960785 issued by JPL/Caltech.

\eject

\clearpage

\begin{deluxetable}{lclclclll}
\tabletypesize{\scriptsize}
\tablecaption{MIPS Measurements of Steep Spectrum Radio Sources}
\tablewidth{0pt}
\tablehead{
\colhead{source}             & \colhead{Integ. time(sec)} & \colhead{24 $\mu$m(mJy)}   &
\colhead{Integ. time(sec)}   & \colhead{70$\mu$m(mJy)}    \\
\colhead{(1)}                & \colhead{(2)}             & \colhead{(3)}              &
\colhead{(4)}                & \colhead{(5)}
 }
\startdata
3C48      & 48.2   & 131.0 $\pm$   0.09       & 37.7      &  696  $\pm$  8       \\
3C61.1    & 48.2   & 5.0   $\pm$   0.07       & 125.8     &  30   $\pm$  4       \\
PicA      &165.7   & 130   $\pm$   0.04       & 125.8     &  171  $\pm$  5       \\
3C171     &165.7   & 7.5   $\pm$   0.05       & 125.8     &  19   $\pm$  3       \\
3C173.1   &165.7   & 0.51  $\pm$   0.06       & 125.8     &  1.7  $\pm$  2       \\
3C218     &165.7   & 8.6   $\pm$   0.06       & 125.8     &  118  $\pm$  5       \\
3C219     &165.7   & 12.7  $\pm$   0.07       & 125.8     &  29   $\pm$  5       \\
3C236     &165.7   & 17.4  $\pm$   0.07       & 125.8     &  50   $\pm$  3       \\
3C249.1   &48.2    & 45.0  $\pm$   0.07       & 125.8     &  57   $\pm$  2       \\
3C277.1   &48.2    & 19.8  $\pm$   0.07       & 125.8     &  20   $\pm$  2       \\
3C284     &165.7   & 24.6  $\pm$   0.07       & 125.8     &  80   $\pm$  2       \\
3C303.1   &165.7   & 7.6   $\pm$   0.05       & 125.8     &  27   $\pm$  15      \\
3C315     & 165.7  & 1.9   $\pm$   0.05       & 125.8     &  28   $\pm$  4       \\
3C323.1   & 48.2   & 33.0  $\pm$   0.08       & 125.8     &  30   $\pm$  2       \\
3C327     &165.7   & 245   $\pm$   0.07       & 125.8     &  468  $\pm$  7       \\
3C348     &165.7   & 0.25  $\pm$   0.07       & 125.8     &  28   $\pm$  5       \\
3C349     &165.7   &       $\le$   0.15       & 125.8     &  7    $\pm$  3       \\
3C351     &48.2    & 103   $\pm$   0.07       & 125.8     &  182  $\pm$  3       \\
3C381     &165.7   & 46    $\pm$   0.05       & 125.8     &  40   $\pm$  3       \\
3C388     &165.7   & 2.34  $\pm$   0.05       & 125.8     &  16   $\pm$  3       \\
\enddata
\tablecomments{Nominal photon-noise errors are given to show the intrinsic signal to noise
ratio for the measurements. Systematic calibration uncertainties are discussed in the text.
The upper limit is at 3$\sigma$ significance. The integration
time at 160$\mu$m was 84 seconds per source. Sources detected at this wavelength are: 
3C48, 1147 $\pm$ 93; PicA, 475 $\pm$ 40 mJy;
3C218, 211 $\pm$ 31 mJy; 3C327, 425 $\pm$ 38 mJy; 3C351, 182 $\pm$ 21 mJy}\\
\end{deluxetable}

\clearpage

\begin{deluxetable}{llllllllllll}
\tabletypesize{\scriptsize}
\tablecaption{Other Properties of the MIPS Sample}
\tablewidth{0pt}
\tablehead{
\colhead{source}                     &\colhead{M$_B$}                &\colhead{Type}                    & \colhead{z}       & \colhead{F$_{5t}$}                &
\colhead{F$_{\rm 5c}$}          &\colhead{Ref}              &\colhead{Size}              & \colhead{$\Gamma_X$}       & \colhead{X-ray}                      &
\colhead{Line}                 &\colhead{Ref}
\\
\colhead {}   & \colhead {} & \colhead {} & \colhead {}   & \colhead {(mJy)}   & \colhead {(mJy)} &
\colhead {} & \colhead {(kpc)}   & \colhead {} & \colhead {}  & \colhead {ratio}  & \colhead {}\\

\colhead {(1)}   & \colhead {(2)} & \colhead {(3)} & \colhead {(4)}   & \colhead {(5)}   & \colhead {(6)} &
\colhead {(7)} & \colhead {(8)}   & \colhead {(9)} & \colhead {(10)}  & \colhead {(11)}  & \colhead {(12)}
}
\startdata
3C48    & -26.5  & Q     & 0.367  &  5,330  $\pm$ 70        &896       &   A91 & 2.4 &  2.12       & 2.39  & -0.013& GW94 \\
3C61.1  & -21.3  & NLRG  & 0.186  & 1,900  $\pm$ 90         &  2.64    &  ZB95 & 535 &             &       &0.47   & La96 \\
PicA    & -19.5  & BLRG  & 0.035  & 15,500 $\pm$ 470        &  1004    &  ZB95 & 281 &  1.69       & 8.66  &0.62   & Ta93 \\
3C171   & -22.0  & NLRG  & 0.238  & 1,210  $\pm$ 60         &  1.68    &  ZB95 &  31 &             &       &0.79   & ST03 \\
3C173.1 & -22.4  & WLRG  & 0.292  & 770    $\pm$ 120        &  7.21    &  ZB95 & 244 &             &       &       &      \\
3C218   & -22.4  & NLRG  & 0.054  & 13,100                  &  245     &  ZB95 &     &  1.79       & 0.224 &-0.125 & Sm96 \\
3C219   & -22.5  & BLRG  & 0.174  & 2,270  $\pm$ 110        &  49.2    &  ZB95 & 441 &  1.58       & 1.62  &0.32   & La96 \\
3C236   & -21.6  & WLRG  & 0.098  & 1,330  $\pm$ 130        &  141     &  ZB95 & 3956&             &       &0.57   & La96 \\
3C249.1 & -26.8  & Q     & 0.311  & 780    $\pm$ 40         &  76      &  K89  & 89  &  1.77       & 1.49  &0.97   & BO84 \\
3C277.1 & -24.7  & Q     & 0.320  & 1,040  $\pm$ 50         &          &       & 6.5 &  1.63       & 0.691 &0.20   & GW94 \\
3C284   & -21.5  & NLRG  & 0.239  & 685    $\pm$ 60         &  3.2     &  Gi88 & 635 &             &       &       &      \\
3C303.1 & -22.0  & NLRG  & 0.267  & 457    $\pm$ 50         &$<$4.7    &  ZB95 & 9.6 &             &       &       &      \\
3C315   & -21.4  & NLRG  & 0.108  & 1,260  $\pm$ 60         &$<$150    &  Gi88 & 257 &             &       &       &      \\
3C323.1 & -24.3  & Q     & 0.264  & 880                     &32        &  K89  & 267 &             &       & 0.71  & BO84 \\
3C327   & -21.9  & NLRG  & 0.103  & 2,740  $\pm$ 140        &  30.4    &  ZB95 & 369 &             &       &1.22   & de92 \\
3C348   & -22.2  & NLRG  & 0.154  & 11,800 $\pm$ 590        &  10.2    &  ZB95 & 286 &             &       &       &      \\
3C349   & -21.1  & NLRG  & 0.205  & 1,130  $\pm$ 60         &  25.0    &  ZB95 & 257 &             &       &       &      \\
3C351   & -27.7  & Q     & 0.371  & 1,200  $\pm$ 60         &  8.0     &  K89  & 277 &  1.69       & 1.36  &-0.016 & CB96 \\
3C381   & -22.0  & BLRG  & 0.160  & 1,280  $\pm$ 50         &  5.12    &  ZB95 & 177 &             &       &1.26   & GO78 \\
3C388   & -23.1  & WLRG  & 0.090  & 1,760  $\pm$ 40         &  59.0    &  ZB95 & 48  &             &       &0.49   & La96 \\
\enddata
\tablecomments{Column (3): The emission-line type, $'$NLRG$'$=Narrow line radio galaxy, $'$BLRG$'$=Broad line radio galaxy,
$'$WLRG$'$=weak line radio galaxy, $'$Q$'$=Quasar. Column (5): The 5 GHz flux density in units of mJy. Column(6): The radio
core flux at 5 GHz in the units of mJy. Column (7): References for Column (6): A91 - Akujor et al. (1991); 
Gi88 - Giovannini et al. (1988); K89 - Kellermann et al. (1989); ZB95- Zirbel \& Baum (1995). Column (8): The linear radio size in
units of kpc. The references for radio sizes are Akujor et al. (1995), Allington-Smith (1984), Gavazzi et al. (1978) and Nilsson et al. (1993). Column (9): The spectral index of hard X-ray emission
over 2-10 kev in the rest frame, assuming a power law SED. Column (10): The rest-frame 2-10 kev
absorption-corrected X-ray flux in the central 2.5 $''$ region. Units are 10$^{-12}$ergs s$^{-1}$ cm$^{-2}$.
Column (11): The logarithm of the ratio of the intensity of [O III]$\lambda$5007 to H$\beta$. Column (12):
References for the Column (11):  BO84 - Boroson \& Oke (1984);
CB96 - Corbin \& Boroson (1996); de92 - de Grijp et al. (1992); Fa89 - Fanti et al. (1989);
Fe97 - Fernini et al. (1997); GO78 - Grandi \& Osterbrock (1978); GW94 - Gelderman \& Whittle (1994);
H89 - Hough et al. (1989); La96 - Lawrence et al. (1996); S87 - Saikia et al. (1987);
Sm96 - Simpson et al. 1996; ST03 - Sol\'orzano-I\~narrea \& Tadhunter (2003); Ta93 - Tadhunter et al. (1993);
V98 - van Bemmel et al. (1998)
}
\end{deluxetable}

\clearpage

\begin{deluxetable}{llllllllllllllll}
\tabletypesize{\scriptsize} \rotate \tablecaption{The Extended
Sample of Radio Galaxies and Quasars} \tablewidth{0pt}
\tablehead{
\colhead{source}          &\colhead{M$_B$}           &\colhead{Type}                          &
\colhead{z}               &\colhead{F$_{5t}$}         &\colhead{F$_{5c}$}                      &
\colhead{Ref}             &\colhead{F$_{24}$}        &\colhead{F$_{70}$}                      &
\colhead{F$_{160}$}       &\colhead{Size}           &\colhead{$\Gamma_X$}                    &
\colhead{X-ray}           &\colhead{Line}            &\colhead{Ref}\\
\colhead {}               &\colhead {}               &\colhead {}                             &
\colhead {}               &\colhead {(mJy)}          &\colhead {(mJy)}                        &
\colhead {}               &\colhead {(mJy)}          &\colhead {(mJy)}                        &
\colhead {(mJy)}          &\colhead {(kpc)}          &\colhead {}                             &
\colhead {}               &\colhead {ratio}          &\colhead {}\\
\colhead {(1)}            &\colhead {(2)}            &\colhead {(3)}                          &
\colhead {(4)}            &\colhead {(5)}            &\colhead {(6)}                          &
\colhead {(7)}            &\colhead {(8)}            &\colhead {(9)}                          &
\colhead {(10)}           &\colhead {(11)}           &\colhead {(12)}                         &
\colhead {(13)}           &\colhead {(14)}           &\colhead {(15)}
 }
\startdata
3C2     & -27.8   & Q    & 1.037 &  1,400  $\pm$ 70         &50        &   S87   &                          & 91 $\pm$  30     &                   &   37&          &     &       &      \\
3C20    & -21.9   & NLRG & 0.174 &  4,150  $\pm$ 210        &2.6       &   Fe97  &                          & 107 $\pm$  35    & 162  $\pm$ 140    &  146&          &     &       &      \\
3C33.1  & -22.2   & BLRG & 0.181 &  860    $\pm$ 40         &14.02     &   ZB95  &   42 $\pm$ 18            & 45  $\pm$  15    &                   &  614&          &     &       &      \\
3C47    & -25.6   & Q    & 0.425 &  1,090  $\pm$ 50         &77.9      &   V98   &   42 $\pm$  30           & 117 $\pm$  37    & 165  $\pm$  71    &  353&1.28      &1.55 & -0.18 & CB96 \\
3C65    & -24.9   & NLRG & 1.176 &  770    $\pm$ 120        &0.50      &   ZB95  &\textbf{24$\pm$193}       & 110 $\pm$  20    & 42   $\pm$  34    &  134&          &     &       &      \\
3C79    & -23.9   & NLRG & 0.255 &  1,300  $\pm$ 70         &10.00     &   ZB95  &   59  $\pm$  15          & 141 $\pm$  26    &                   &  322&          &     & 0.40  & Sc65 \\
3C109   & -25.7   & Q    & 0.305 &  1,630  $\pm$ 160        &305.85    &   ZB95  &   165 $\pm$  41          & 208 $\pm$  34    &                   &  403&1.69      &8.66 &       &      \\
3C111   & -23.8   & Q    & 0.049 &  7,870  $\pm$ 393        &          &         &   213 $\pm$  54          & 342 $\pm$  100   &                   &  244&          &     &       &      \\
3C234   & -22.3   & NLRG & 0.184 &  1,530  $\pm$ 50         &95.79     &   ZB95  &   259 $\pm$  75          & 269 $\pm$  54    &                   &  317&          &     & 0.48  & Sc65 \\
3C277.2 & -23.9   & NLRG & 0.766 &  480                     &$<$15.04  &   ZB95  &                          & 122 $\pm$  42    &                   &  379&          &     &       &      \\
3C293   & -21.5   & WLRG & 0.045 &  1,857  $\pm$ 40         & 100      &   Gi88  &   42  $\pm$  6           & 305 $\pm$  31    &                   &  70 &          &     &       &      \\
3C298   & -32.2   & Q    & 1.439 &  1,450  $\pm$ 70         &          &         &   37  $\pm$  12          & 191 $\pm$  27    & 238  $\pm$  74    &   15&1.75      &2.25 &       &      \\
3C309.1 & -29.5   & Q    & 0.904 &  3,730  $\pm$ 190        &2350      &   H89   &\textbf{36$\pm$133}       & 114 $\pm$  13    & 251  $\pm$  84    &   15&1.40      &1.44 & 0.14  & La96 \\
3C318   & -25.1   & Q    & 0.752 &  750                     &$<$44.43  &   ZB95  &                          & 197 $\pm$  64    & 379  $\pm$ 114    &  6.8&          &     &       &      \\
3C321   & -22.3   & NLRG & 0.0961&  1,210  $\pm$ 120        &30        &   Gi88  &   326  $\pm$ 6           & 1006 $\pm$ 46    &                   &  482&          &     &       &      \\
3C325   & -24.9   & Q    & 0.860 &  830    $\pm$ 120        &2.4       &   Fe97  &\textbf{27$\pm$63}        & 154 $\pm$  52    &\textbf{35$\pm$46} &  114&          &     &       &      \\
3C332   & -23.2   & Q    & 0.1515&  830    $\pm$ 42         &          &         &   18  $\pm$  5           & 52  $\pm$  30    &                   & 224 &          &     &       &      \\ 
3C334   & -28.2   & Q    & 0.555 &  620                     &159.5     &   V98   &                          & 75  $\pm$  21    & 55   $\pm$  24    &  276&1.59      &0.935&       &      \\
3C368   & -26.5   & NLRG & 1.132 &  210    $\pm$ 30         &$<$0.72   &   ZB95  &\textbf{39$\pm$77}        & 87 $\pm$  28     &\textbf{79 $\pm$82}&   60&          &     &       &      \\
3C380   & -28.2   & Q    & 0.691 &  7,450  $\pm$ 370        &7447      &   H89   &   13  $\pm$   7.5        & 55  $\pm$  17    & 78   $\pm$  33    &   49&1.32      &2.48 & 0.15  & GW94 \\
3C382   & -23.1   & Q    & 0.058 &  2,220  $\pm$ 111        &188       &   Gi88  &   93  $\pm$   23         & 107 $\pm$  31    &                   &  188&          &     &       &      \\ 
3C390.3 & -20.8   & BLRG & 0.056 &  4,450  $\pm$ 79         &344.37    &   ZB95  &   320 $\pm$   7.6        & 237 $\pm$   6    &                   &  208&          &     & -0.24 & La96 \\
3C403   & -21.7   & NLRG & 0.059 &  2,060  $\pm$ 100        &          &         &   220 $\pm$   30         & 503 $\pm$  52    &                   & 103 &          &     &       &      \\ 
3C405   & -21.7   & NLRG & 0.056 &  371,000$\pm$ 18,500       &314.07    &   ZB95  &   834 $\pm$ 113          & 2771 $\pm$  56   & 430  $\pm$ 230    &  123&0.89      &18.5 & 1.11  & Ta94 \\
3C433   & -21.7   & NLRG & 0.102 &  3,710  $\pm$ 190        &5         &   Gi88  &   193 $\pm$ 48           & 319  $\pm$ 93    &                   & 87  &          &     &       &      \\
3C445   & -20.6   & BLRG & 0.056 &  2,030  $\pm$ 100        &88.52     &   ZB95  &   315 $\pm$  47          &                  &                   &  730&          &     & 1.11  & Ta93 \\
3C459   & -23.0   & NLRG & 0.219 &  1,350  $\pm$ 70         &1138.26   &   ZB95  &                          &962 $\pm$ 176     & 947  $\pm$ 351    &   27&          &     &       &      \\
\enddata
\tablecomments{Column (3): The emission-line type, $'$NLRG$'$=Narrow line radio galaxy, $'$BLRG$'$=Broad line radio
galaxy, $'$WLRG$'$=weak line radio galaxy, $'$Q$'$=Quasar. Column (5): The 5 GHz flux density in units of mJy.
Column(6): The radio core flux at 5 GHz in units of mJy. Column (7): References for the Column (6):
Fe97 - Fernini et al. (1997); Gi88 - Giovannini et al. (1988); H89 - Hough et al. (1989); S87 - Saikia et al. (1987); 
ZB95- Zirbel \& Baum (1995); V98 - van Bemmel et al. (1998).
Column (8), (9) and (10): The flux with uncertainty at 24 $\mu$m, 70 $\mu$m, 160 $\mu$m, respectively.
Bold indicates the sources are not detected to a 3-sigma level.  Column (11): The linear
radio size in units of kpc. The references for radio sizes are Akujor et al. (1995), Allington-Smith (1984), Gavazzi et al. (1978) and Nilsson et al. (1993).
Column (12): The spectral index of the hard X-ray emission at 2-10 kev in the rest frame,
assuming a power law SED. Column (13): The rest-frame 2-10 kev absorption-corrected X-ray flux in the
central 2.5 $''$ region. Units are 10$^{-12}$ergs s$^{-1}$ cm$^{-2}$. Column (14): The logarithm of the ratio
of the intensity of [O III]$\lambda$5007 to H$\beta$. Column (15): References for Column (14):
CB96 - Corbin \& Boroson (1996); Fa89 - Fanti et al. (1989); GW94 - Gelderman \& Whittle (1994); La96 - Lawrence et al. (1996); Sc65 - Schmidt (1965); Ta93 - Tadhunter et al. (1993);
Ta94 - Tadhunter et al. (1994).}

\end{deluxetable}


\clearpage

\begin{deluxetable}{lllllll}
\tabletypesize{\scriptsize}
\tablecaption{Kendall Tau test}
\tablewidth{0pt}
\tablehead{
\colhead{relation} & \colhead{$r_{\rm lum}$}         & \colhead{$S_{\rm lum}$} & \colhead{$r_{\rm flux}$} & \colhead{$S_{\rm flux}$}
}
\startdata

X-ray      vs 70 $\mu$m     & --   &  --       & 0.32     &   0.09  \\
X-ray      vs 24 $\mu$m     & --   &  --       & 0.37     &   0.05  \\
X-ray      vs radio core    & --    & --       & 0.30     &   0.14  \\
IR color   vs $R$ parameter & --    & --       & -0.37    &   0.04  \\
178MHz vs 70 $\mu$m (all)  & --    &  --       & 0.25      &   0.02  \\
178MHz vs 70 $\mu$m (quasars)& --  &  --       & 0.68      &   0.0002\\
Radio size  vs 24  $\mu$m  & -0.01 &  0.96    & --        &  --      \\
Radio size  vs 70  $\mu$m  & -0.11 &  0.52    & --        &  --     \\
\enddata
\tablecomments{The $r$ value indicates the rank correlation coefficient
and the $S$ value indicates the two-sided significance of
the deviation from zero.}
\end{deluxetable}


\clearpage

\begin{figure}
\epsscale{0.80}
\plotone{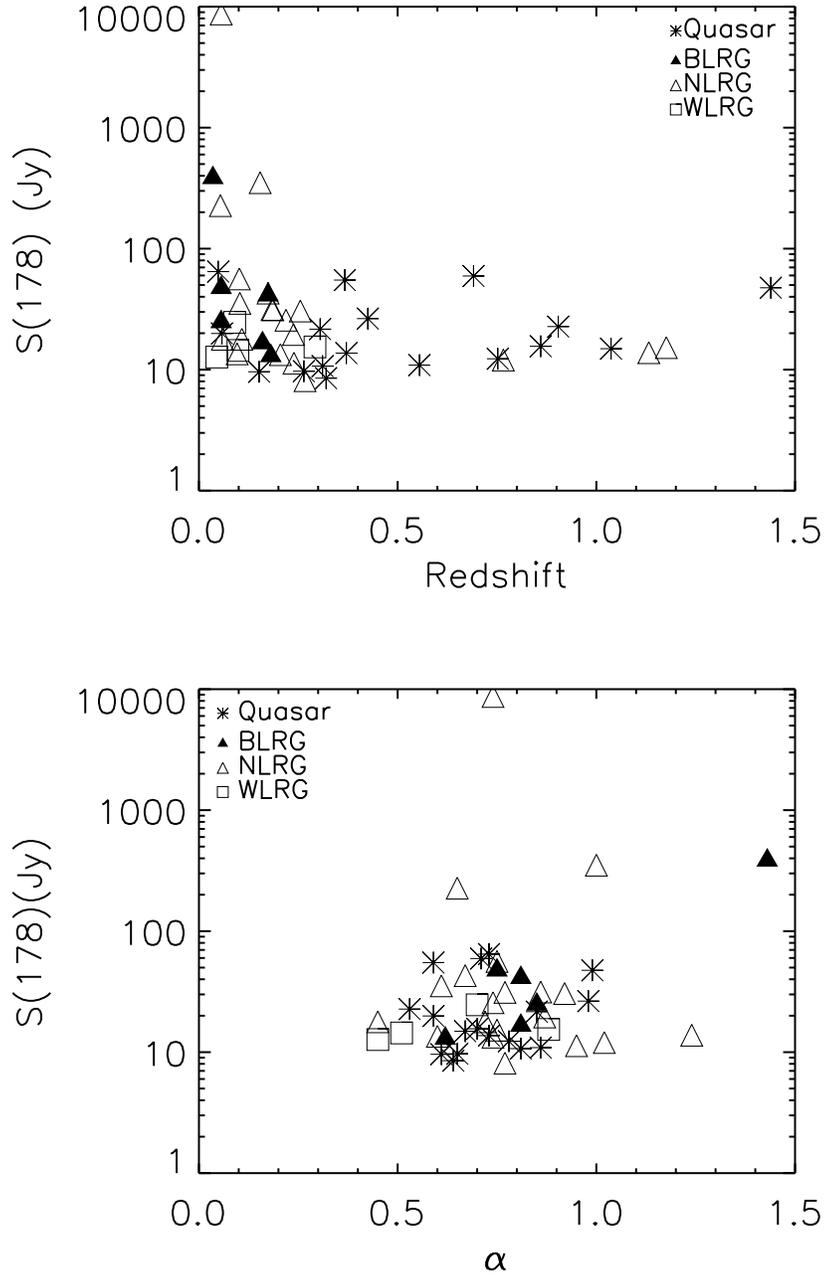}
\caption{The distributions of
sources in the plane of the 178 MHz flux versus redshift (top
panel) and 178 MHz flux versus spectral index $\alpha$ (bottom
panel).}
\end{figure}

\clearpage

\setcounter{figure}{1}
\begin{figure}
\epsscale{1.0}
\plotone{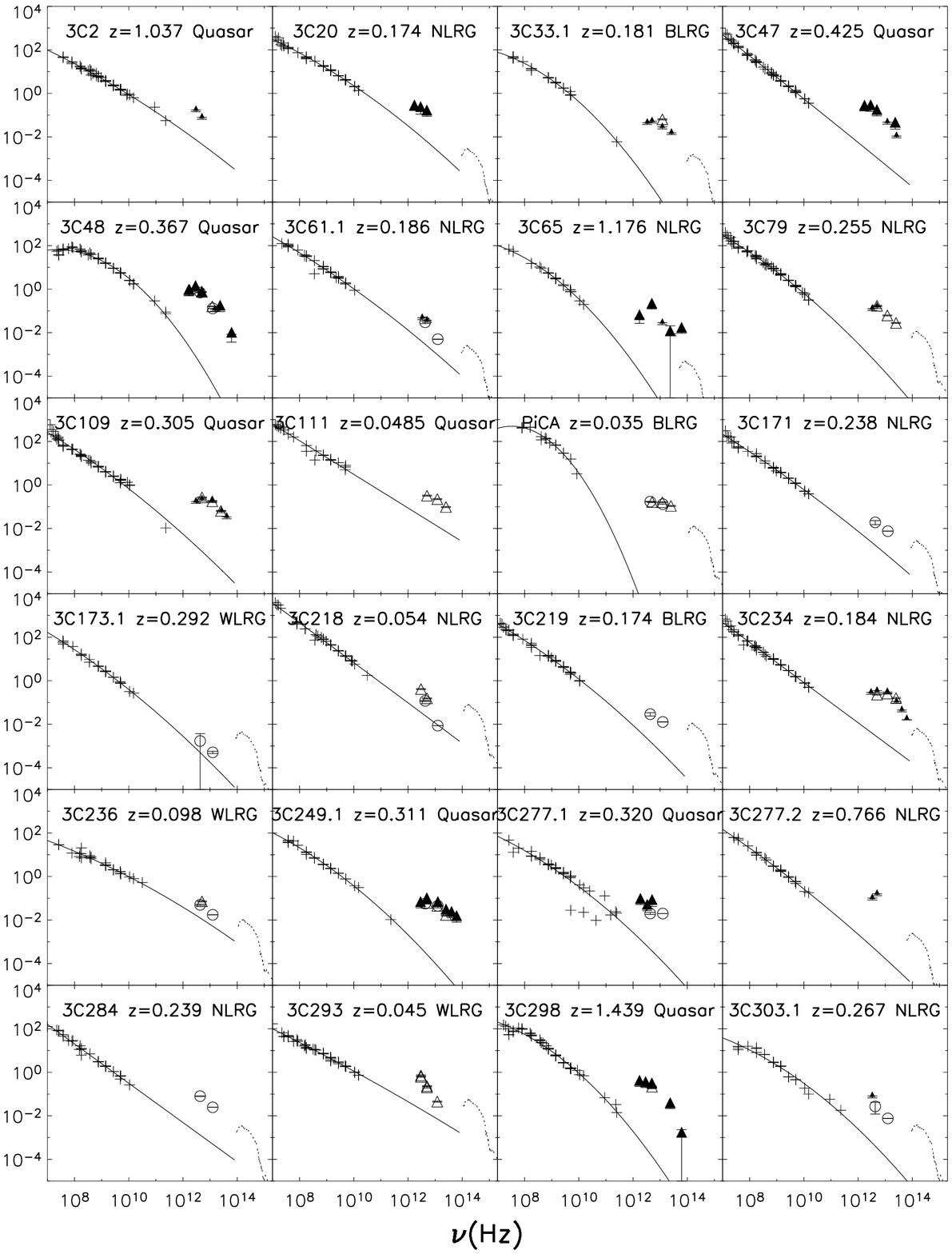}
\caption{}
\end{figure}

\setcounter{figure}{1}
\begin{figure}
\epsscale{1.0}
\plotone{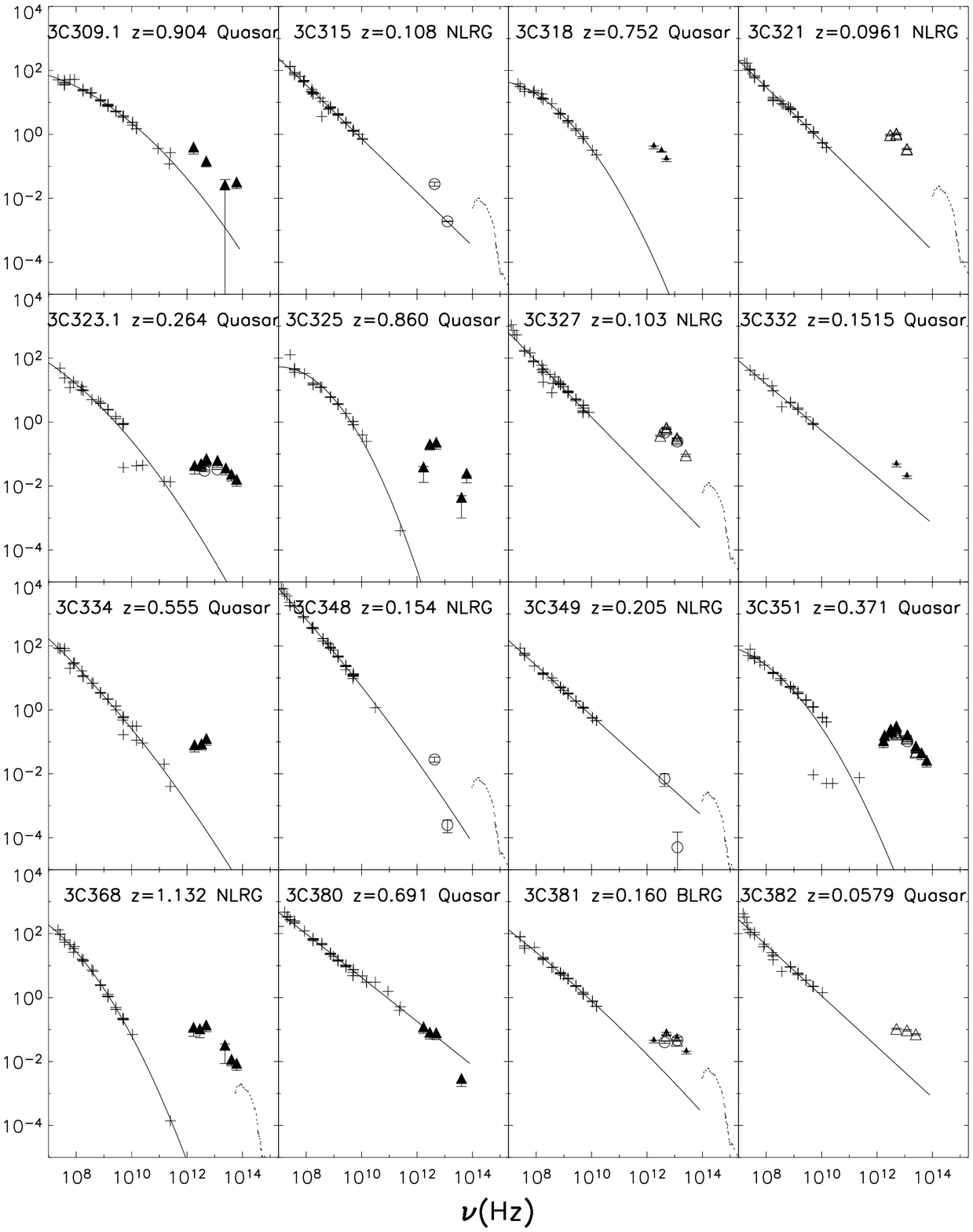}
\caption{}
\end{figure}

\setcounter{figure}{1}
\begin{figure}
\epsscale{.80}
\plotone{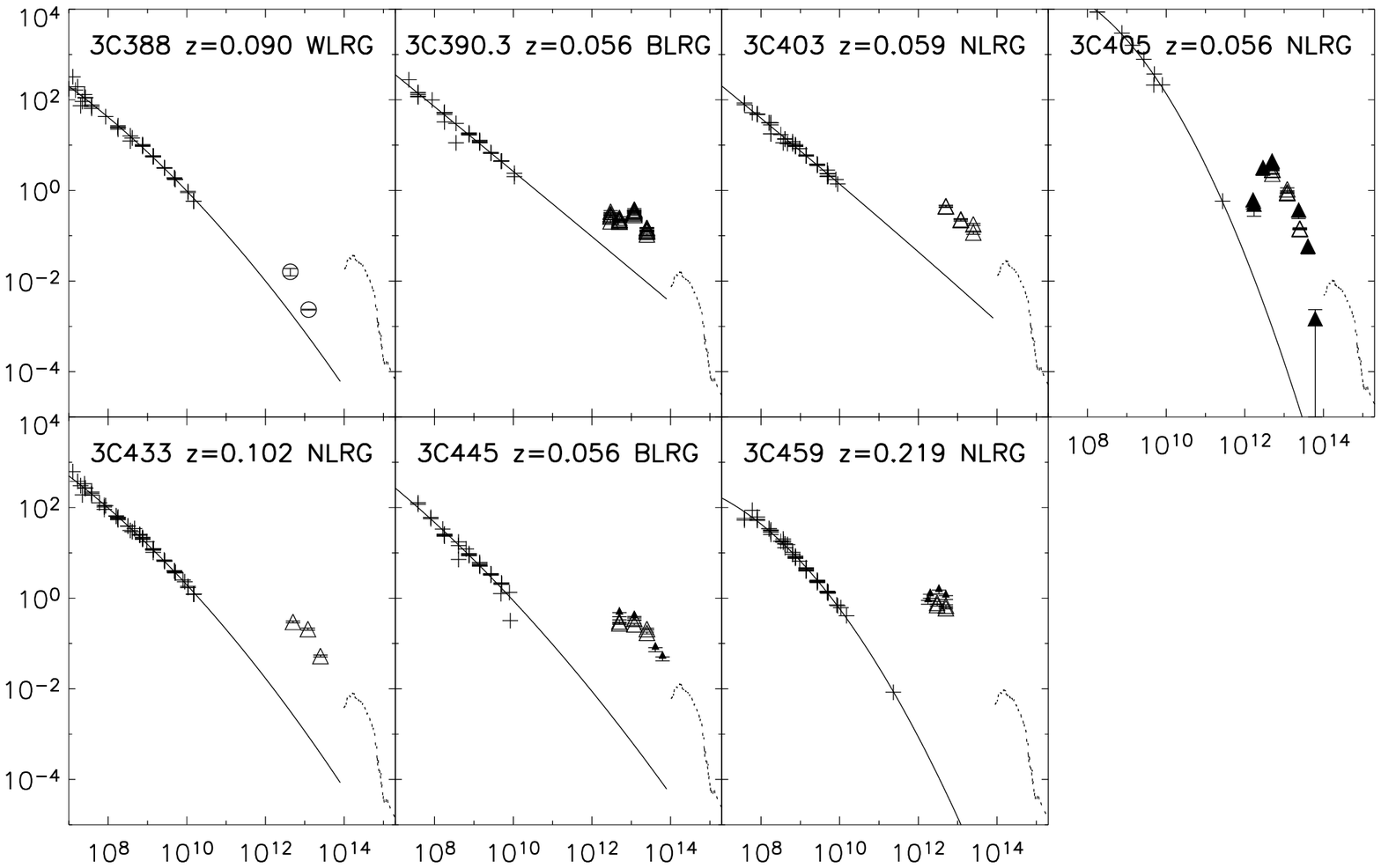} 
\caption{The
spectral energy distributions of all sources. The 1-sigma error is used in the plot. The crosses are
the radio and millimeter photometry data from NED. The solid line
is the parabola model for the synchrotron emission. The open
circle, open triangle, and filled triangle denote MIPS, $IRAS$ and
$ISO$ data, respectively. The dotted line for radio galaxies
is the template of stellar light for a normal elliptical galaxy,
normalized to the apparent B magnitude. $'$Quasar$'$, $'$BLRG$'$, $'$NLRG$'$ and $'$WLRG$'$ indicate
Quasar, broad line radio galaxies, narrow line radio galaxies and
weak line radio galaxies, respectively.}
\end{figure}

\clearpage

\begin{figure}
\epsscale{0.8}
\plotone{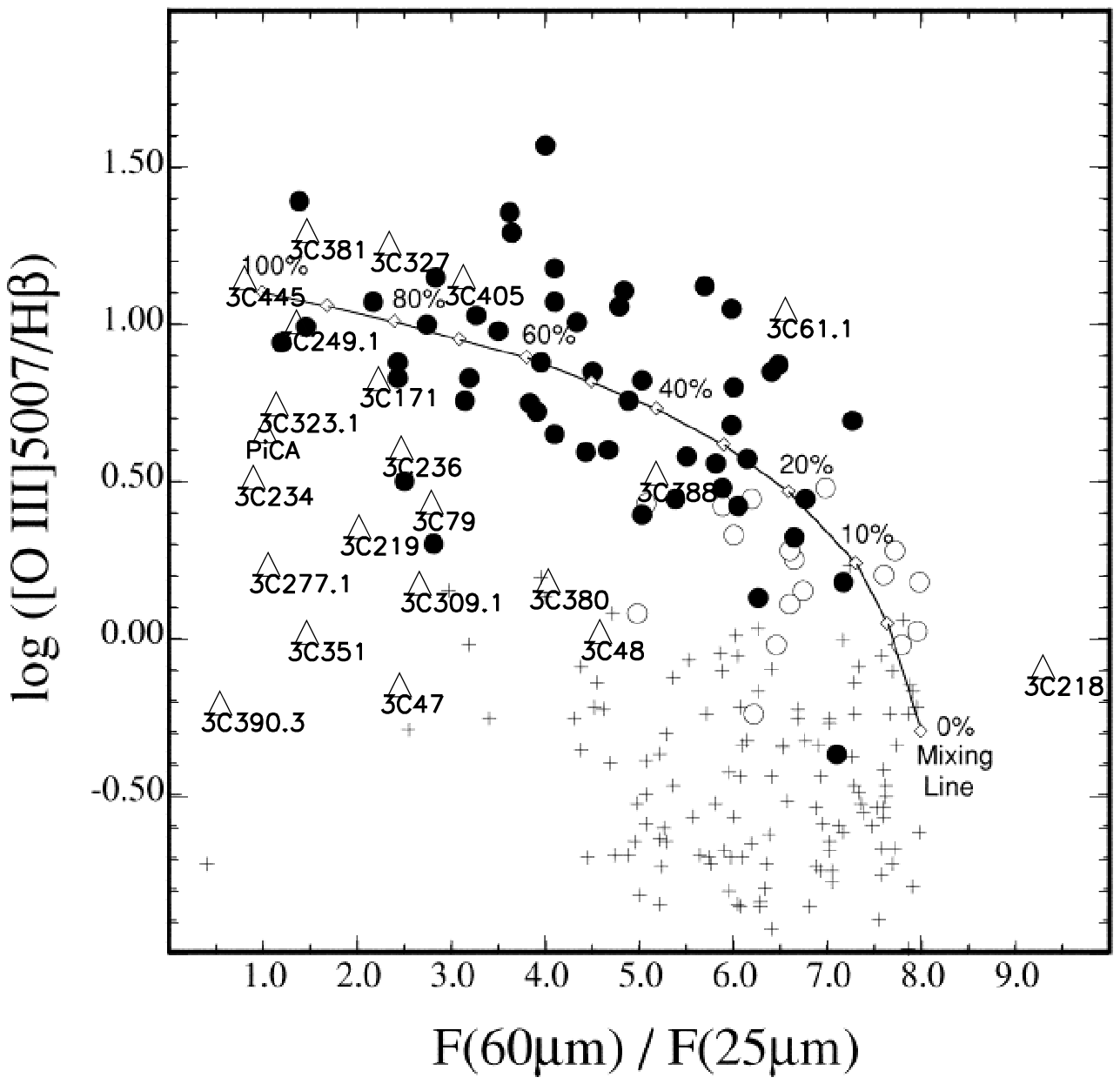} 
\caption{IR color
versus emission-line ratio. The dashed line is a hypothetical
mixing line on which the AGN fraction is labelled. This figure is from
Kewley et al. (2001), with our sources added. The open triangles denote our sources. 
Filled circles, unfilled circles and crosses denote 
AGNs, ambiguous classification and starbursts, respectively in Kewley et al. (2001). 
Note that Section 4.2.1 shows that the torus may have significant optical depth 
at 24 $\mu$m and thus the IR color underestimates the fraction of AGN contributions for Type II AGN.}
\end{figure}

\clearpage

\begin{figure}
\epsscale{1.0}
\plotone{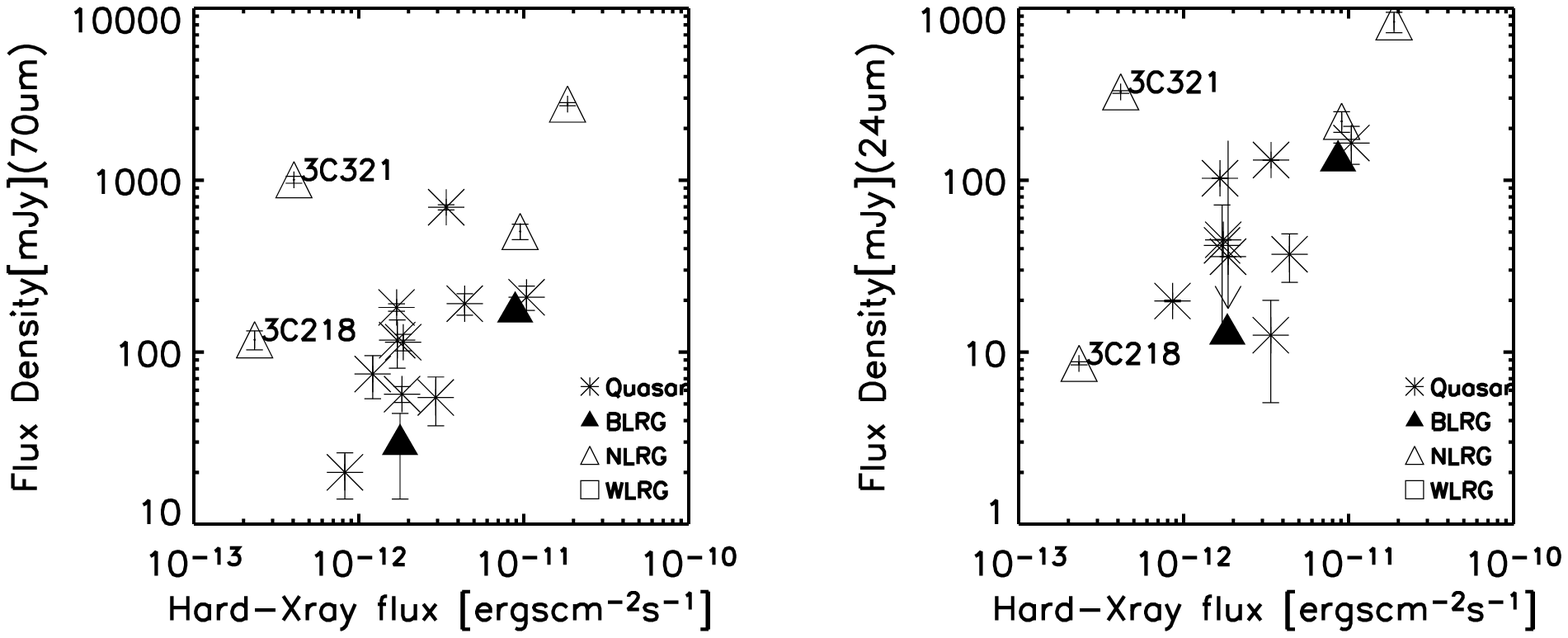} 
\caption{The
correlation of K-corrected 70 $\mu$m flux density with 1-sigma error, K-corrected 24 $\mu$m flux density with 1-sigma error
and central K-corrected hard X-ray flux corrected for absorption.
}
\end{figure}

\clearpage

\begin{figure}
\epsscale{0.8}
\plotone{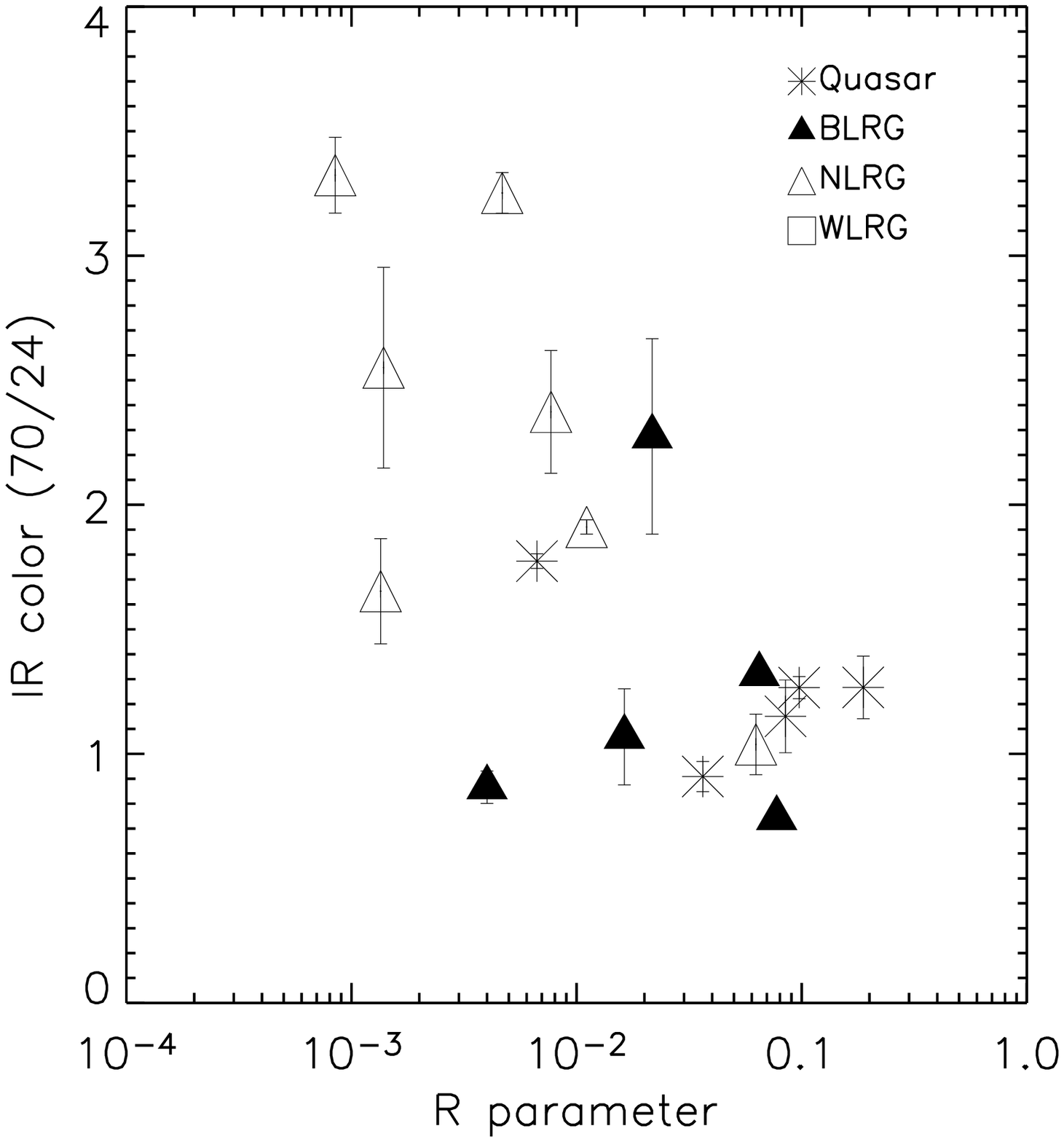} 
\caption{The
correlation between IR color with 1-sigma error and $R$ parameter. The IR color is defined as the ratio of 70 $\mu$m flux
to 24 $\mu$m flux. The $R$ parameter is defined as the ratio of the core radio flux to the total radio flux at 5 GHz. The solid line shows the least square fit.
 }
\end{figure}

\clearpage

\begin{figure}
\epsscale{0.8}
\plotone{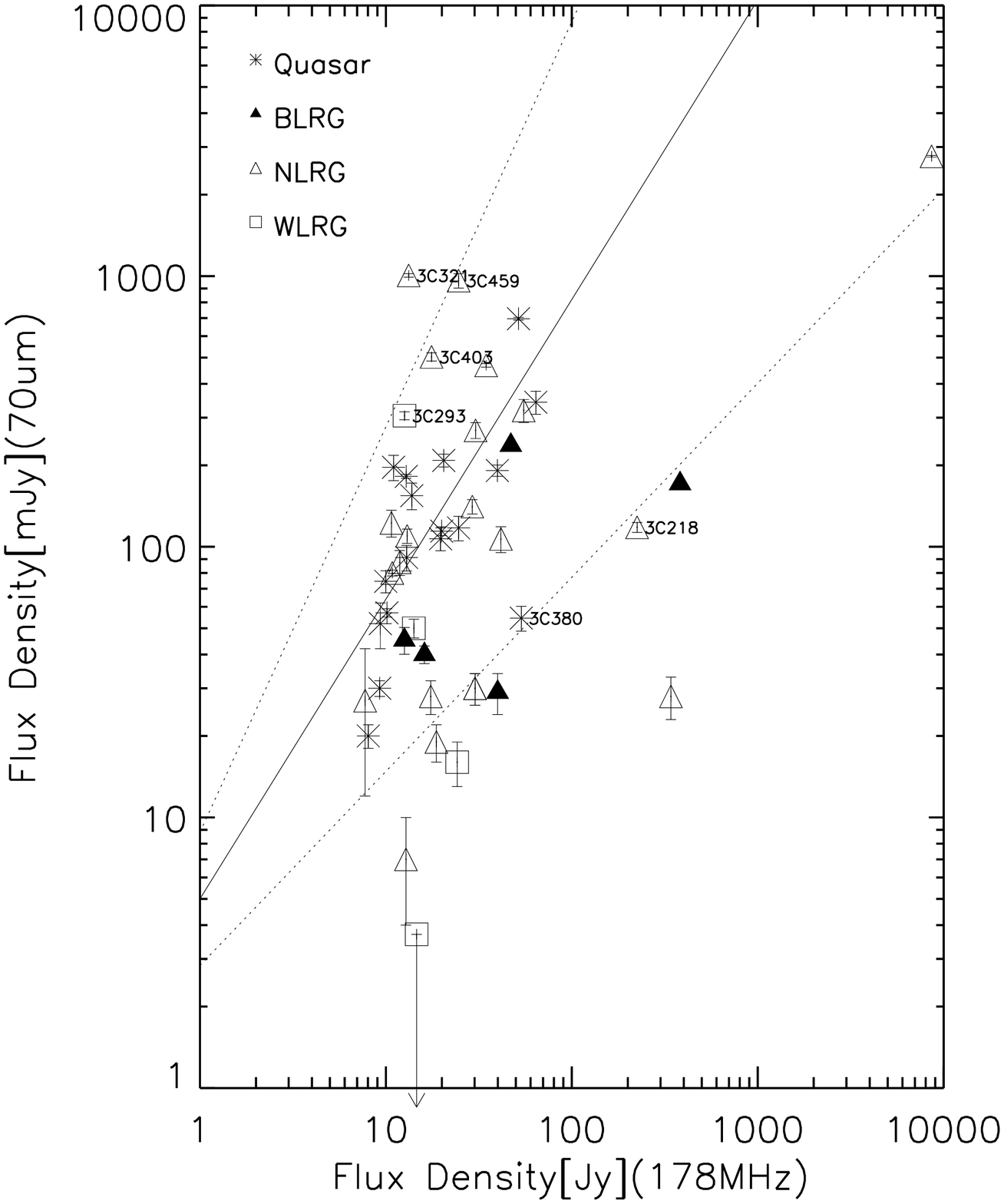} 
\caption{The
relation of the K-corrected 70 $\mu$m emission with 1-sigma error and K-corrected radio emission at 178 MHz.
The solid line is the least square fit to the radio
quasars, excluding 3C 380 with substantial contribution by non-thermal core ouput (See Section 3.2), unlike those of the other quasars.
The two dashed lines are the 3-sigma scatter of the relation indicated by the solid line.
}
\end{figure}

\clearpage

\begin{figure}
\epsscale{0.8}
\plotone{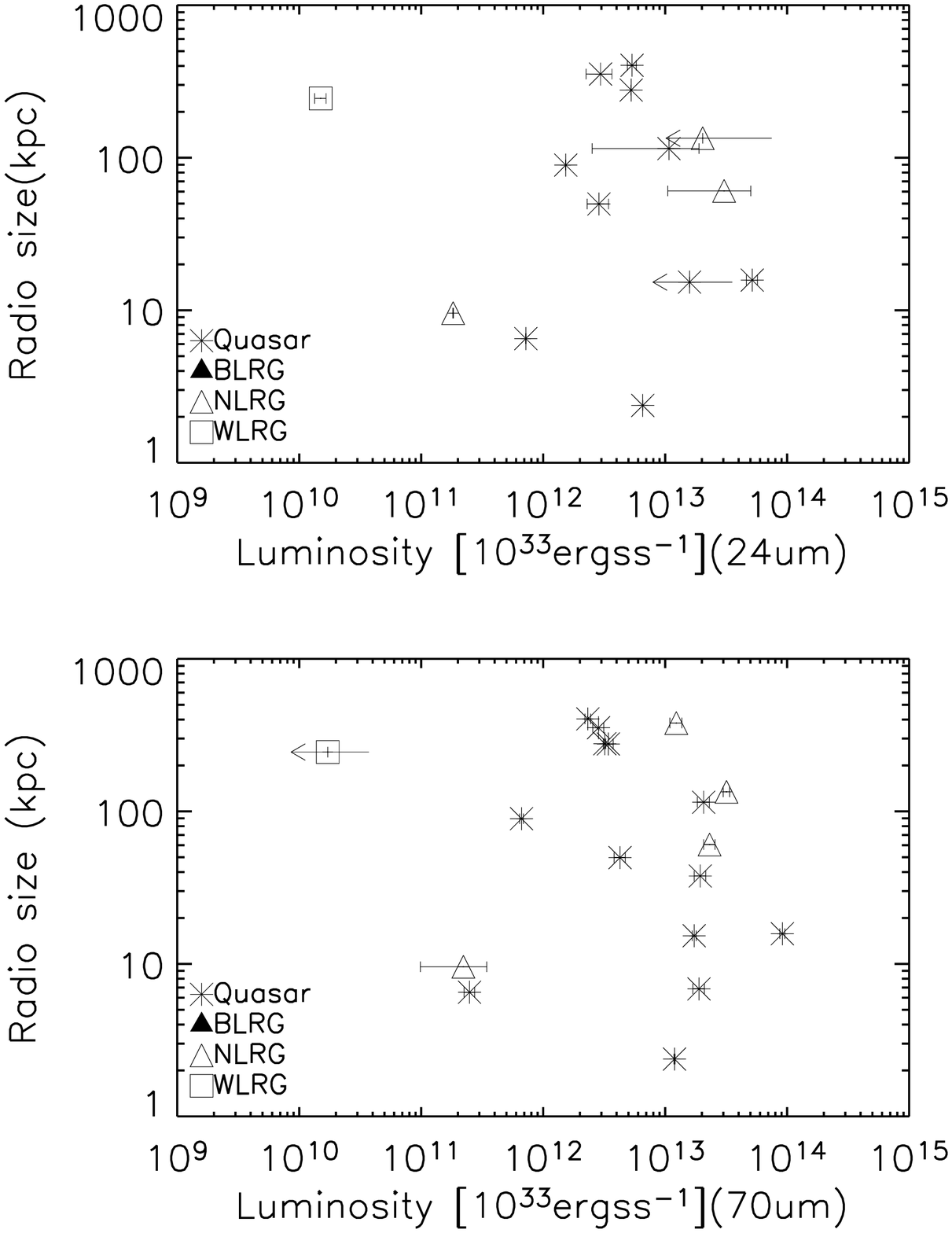} 
\caption{The plot of linear radio size vs. K-corrected FIR luminosity with 1-sigma error.}
\end{figure}


\begin{thebibliography}{}
\bibitem[AK1994]{629} Akujor, C. E. \& Garrington, S. T. 1995, A\&AS, 112, 235

\bibitem[Ak1991]{631} Akujor, C. E., Spencer, R. E., Zhang, F. J., Davis, R. J., Browne, I. W. A. \& Fanti, C. 1991, MNRAS, 250, 215

\bibitem[Al1984]{633} Allington-Smith, J. R. 1984, MNRAS, 210, 611

\bibitem[An2002]{635} Andreani, P., Fosbury, R. A., E., van Bemmel, I., \& Freudling, W. 2002, A\&A, 381, 389

\bibitem[Ba1989]{637} Barthel, P. D., 1989, ApJ, 336, 606

\bibitem[Ba1996]{   } Barthel, P. D. \& Arnaud, K. A. 1996, MNRAS, 283, L45

\bibitem[Ba2002]{639} Baker, J. C., Hunstead, R. W., Athreya, R. M., Barthel, P. D., de Silva, E., Lehnert, M. D. \& SaundersR. D. E. 2002, ApJ, 568, 592

\bibitem[Ba1981]{641} Baldwin, J. A., Phillips, M. M. \& Terlevich, R. 1981, PASP, 93, 5

\bibitem[Bi1995]{643} Biretta, J. A., Burrows, C. J., Holtzman, J. A., et al., 1996, In: Biretta J. A. (ed.) Wide Field and Planetary Camera 2 Instrument Handbook. STScI, Baltimore

\bibitem[Bl1994]{   } Bloom, S. D., Marscher, A. P., Gear, W. K., Terasranta, H., Valtaoja, E., Aller, H. D. \& Aller, M. F. 1994, AJ, 108, 398

\bibitem[Bo1984]{645} Boroson, T. A. \& Oke, J. B., 1984, ApJ, 281, 535

\bibitem[Ch2000]{647} Chiaberge, M., Capetti, A., Celotti, A. 2000, A\&A, 355, 873

\bibitem[Co1996]{649} Corbin, M. R. \& Boroson, T. A. 1996, ApJS, 107, 69

\bibitem[de1992]{651} de Grijp, M. H., K., Keel, W. C., Miley, G. K., Goudfrooij, P. \& Lub, J. 1992, A\&AS, 96, 389

\bibitem[de1991]{653} de Vaucouleurs, G., de Vaucouleurs, A., Corwin, H. G., Jr., Buta, R. J., Paturel, G. \& Fouque, P. 1991, S\&T, 82Q, 621D

\bibitem[Fa1989]{655} Fanti, C., Fanti, R., Parma, P., Venturi, T., Schilizzi, R. T., et al. 1989, A\&A, 217, 44

\bibitem[Fa2000]{657} Fanti, C., Pozzi, F., Fanti, R., et al., 2000, A\&A, 358, 499

\bibitem[Fe1997]{659} Fernini, I., Burns, J. O. \& Perley, R. A. 1997, AJ, 114, 2292

\bibitem[FR1974]{661} Fanaroff, B. L. \& Riley, J. M. 1974, MNRAS, 167, 31

\bibitem[Ga1978]{663} Gavazzi, G., Perola, G. C., 1978, A\&A, 66, 407

\bibitem[Ge1994]{665} Gelderman, R. \& Whittle, M. 1994, ApJS, 91, 491

\bibitem[Gi1988]{667} Giovannini, G., Feretti L., Gregorini L. \& Parma P. 1988, A\&A, 199, 73

\bibitem[Go1988]{669} Golombek, D., Miley, G. K., \& Neugebauer, G. 1988, 1988, AJ, 95, 26

\bibitem[G2004a]{671} Gordon, K. D., et al. 2004a, PASP, submitted

\bibitem[G2004b]{673} Gordon, K. D., et al. 2004b, Proc. SPIE, in press

\bibitem[Gr1994]{675} Granato G. L. \& Danese, L. 1994, MNRAS, 268, 235

\bibitem[Gr1978]{677} Grandi, S. A. \& Osterbrock, D. E. 1978, ApJ, 220, 783

\bibitem[Ha1998]{679} Haas, M., Chini, R., Meisenheimer, K., Stickel, M., Lemke, D., Klaas, U., Kreysa, E. 1998, ApJL, 503, L109

\bibitem[Ha2004]{681} Haas, M., M\"ueller, S. A., H., Bertoldi, F., Chini, R., Egner, S., Freudling, W., Klaas, U., Krause, O., Lemke, D., Meisenheimer, K., Siebenmorgen, R., van Bemmel, I. 2004, astoph/0406111

\bibitem[He1994]{683} Heckman, T. M., O'Dea, C. P., Baum, S. A., \& Laurikainen, E. 1994, ApJ, 428, 65

\bibitem[He1995]{685} Hes, R., Barthel, P. D. \& Hoekstra, H. 1995, A\&A, 303, 8

\bibitem[Ho1989]{687} Hough, D. H. \& Readhead, A. C. S. 1989, AJ, 98, 1208

\bibitem[Ke1989]{689} Kellermann, K. I., Sramek, R., Schmidt, M., Shaffer, D. B. \& Green, R. 1989, AJ, 98, 1195

\bibitem[Ke2001]{691} Kewley, L. J., Heisler, C. A., Dopita, M. A. \& Lumsden, S. 2001, ApJS, 132, 37

\bibitem[La1996]{693} Lawrence, C. R., Zucker, J. R., Readhead, A. C. S., Unwin, S. C., Pearson, T. J. \& Xu, W. 1996, ApJS, 107, 541

\bibitem[Lo1987]{695} Lonsdale, C. J. \& Barthel, P. D. 1987, AJ, 94, 1487

\bibitem[Ne2002]{697} Nenkova, M., Ivezic, Z. \& Elitzur, M 2002, ApJ, 570, 9

\bibitem[Ne1986]{699} Neugebauer, G., Miley, G. K., Soifer, B. T., \& Clegg, P. E. 1986, ApJ, 308, 815

\bibitem[Ni1993]{701} Nilsson, K., Valtonen, M. J., Kotilainen, J., Jaakkola, T., 1993, ApJ, 413, 453

\bibitem[Od1998]{703} O'Dea, C. P. 1998, PASP, 110, 493

\bibitem[Or1982]{705} Orr, M. J. L. \& Browne, I. W. A. 1982, MNRAS, 200, 1067

\bibitem[Os1985]{707} Osterbrock, D. E. \& de Robertis, M. M. 1985, PASP, 97, 1129

\bibitem[Pe1976]{709} Pence, W., 1976, ApJ, 203, 39

\bibitem[Pi1993]{711} Pier, A. P. \& Krolik, J. H. 1993, ApJ, 418, 673

\bibitem[Pi2003]{713} Pihlstr\"om, Y. M., Conway, J. E., \& Vermeulen, R. C. 2003, A\&A, 404, 871

\bibitem[Po1998]{   } Polatidis, A. G. \& Wilkinson, P. N. 1998, MNRAS, 294, 327

\bibitem[Po2000]{715} Polletta, M., Courvoisier, T. J., L., Hooper, E. J., \& Wilkes, B. J. 2000, A\&A, 362, 715

\bibitem[Ri2004]{717} Rieke, G. H., Young, E. T., Engelbracht, C., et al. 2004, ApJS, 154, 25

\bibitem[Ro2000]{   } Roche, N., Eales, S. A. 2000, MNRAS, 317, 120

\bibitem[Ro2000]{719} Rowan-Robinson, M. 2000, MNRAS, 316, 885

\bibitem[Sa1987]{721} Saikia, D. J., Salter, C. J. \& Muxlow, T. W., B. 1987, MNRAS, 224, 911

\bibitem[Sa1965]{723} Sandage, A.; V\'eron, P. \& Wyndham, J. D. 1965, ApJ, 142, 1307

\bibitem[Sa1988]{   } Sanders, D. B., Soifer, B. T., Elias, J. H., Madore, B. F., Matthews, K., Neugebauer, G., Scoville, N. Z. 1988, ApJ, 325, 74

\bibitem[Sa1979]{725} Savage, B. D. \& Mathis, J. S. 1979, ARA\&A, 17, 73

\bibitem[Sc1965]{727} Schmidt, M. 1965, ApJ, 141, 1

\bibitem[Sc2002]{729} Schneider, G. \& Stobie, E. 2002, ASP Conf. Ser. 281, Astronomical Data Analysis Software and System XI, ed. D. A. Bohlender , D. Durand \& T. H. Handley (San Francisco: ASP), p. 382

\bibitem[Sm1989]{731} Smith, E. P. \& Heckman, T. M., 1989, ApJ, 69, 365

\bibitem[So2003]{733} Sol\'orzano-I\~narrea, C. \& Tadhunter, C. N. 2003, MNRAS, 340, 705

\bibitem[Sp1985]{735} Spinrad, H., Marr, J., Aguilar, L. \& Djorgovski, S. 1985, PASP, 97, 932

\bibitem[Si2004]{737} Siebenmorgen, R., Freudling, W., Kr\"ugel, E. \& Haas, M. 2004, A\&A, 421, 129

\bibitem[Sm1996]{739} Simpson, C., Ward, M., Clements, D. L., Rawlings, S. 1996, MNRAS, 281, 509

\bibitem[Ta1993]{741} Tadhunter, C. N., Morganti, R., di Serego-Alighieri, S., Fosbury, R. A. E. \& Danziger, I. J. 1993, MNRAS, 263, 999

\bibitem[Ta1994]{743} Tadhunter, C. N., Metz, S. \& Robinson, A. 1994, MNRAS, 268, 989

\bibitem[Ta1996]{   } Tadhunter, C. N., Dickson, R. C., Shaw, M. A. 1996, MNRAS, 281, 591

\bibitem[UP1995]{745} Urry, C. M., \& Padovani, P. 1995, PASP, 107, 803

\bibitem[va1998]{747} van Bemmel, I. M., Barthel, P. D. \& Yun, M. S. 1998, A\&A, 334, 799

\bibitem[va2001]{   } van Bemmel, I., Barthel, P. 2001, A\&A, 379, L21

\bibitem[va2003]{   } van Bemmel, I., Dullemond, C. P. 2003, A\&A, 404, 1

\bibitem[Ve1985]{749} Veilleux, S. \& Osterbrock, D. E. 1987, ApJS, 63, 295

\bibitem[Wi1990]{   } Wilkinson P. N., Tzioumis A. K., Akujor C. E., Benson J. M., Walker R. C., Simon R. S. 1990, in Zensus J. A., Pearson T. J.,
eds, Parsec Scale Radio-Jets. Cambridge Univ. Press, Cambridge, p. 152

\bibitem[Wi2000]{751} Willott, C. J., Rawlings, S., Blundell, K. M. \& Lacy, M. 2000, MNRAS, 316, 449

\bibitem[Wh2004]{753} Whysong, D. \& Antonucci, R. 2004, ApJ, 602, 116

\bibitem[Zi1995]{755} Zirbel, E. L. \& Baum, S. A. 1995, ApJ, 448, 521
\end{thebibliography}
\end{document}